# User Experience Evaluation of Augmented Reality: A Systematic Literature Review

Stefan Graser
stefan.graser@hs-rm.de
*Center for Advanced E-Business Studies*
*RheinMain University of Applied Sciences*
Wiesbaden, Germany
ORCID: 0000-0002-5221-2959

Felix Kirschenlohr

Schwarz IT KG
Neckarsulm, Germany

Stephan Böhm
stephan.boehm@hs-rm.de
*Center for Advanced E-Business Studies*
*RheinMain University of Applied Sciences*
Wiesbaden, Germany
ORCID: 0000-0003-3580-1038

*Abstract*—Due to technological development, Augmented Reality (AR) can be applied in different domains. However, innovative technologies refer to new interaction paradigms, thus creating a new experience for the user. This so-called User Experience (UX) is essential for developing and designing interactive products. Moreover, UX must be measured to get insights into the user's perception and, thus, to improve innovative technologies. We conducted a Systematic Literature Review (SLR) to provide an overview of the current research concerning UX evaluation of AR. In particular, we aim to identify (1) research referring to UX evaluation of AR and (2) articles containing AR-specific UX models or frameworks concerning the theoretical foundation. The SLR is a five-step approach including five scopes. From a total of 498 records based on eight search terms referring to two databases, 30 relevant articles were identified and further analyzed. Results show that most approaches concerning UX evaluation of AR are quantitative. In summary, five UX models/frameworks were identified. Concerning the UX evaluation results of AR in Training and Education, the UX was consistently positive. Negative aspects refer to errors and deficiencies concerning the AR system and its functionality. No specific metric for UX evaluation of AR in the field of Training and Education exists. Only three AR-specific standardized UX questionnaires could be found. However, the questionnaires do not refer to the field of Training and Education. Thus, there is a lack of research in the field of UX evaluation of AR in Training and Education.

## I. Introduction

Over the last decades, AR as an innovative technology has emerged in different domains. AR, therefore, enhances the real environment with digital information and 3D data using different devices [1]. To separate AR from other technologies, [1] defined three characteristics of the technology: Therefore, AR (1) combines reality with virtuality, (2) creates an interaction of both in real-time, and (3) registers digital content in 3-D [1]. Due to technological development, AR can be easily deployed into application fields of daily life, such as Education, Entertainment, or Medicine [2]. Concerning the field of education, AR provides the potential for improving teaching as well as learning [3][4]. The field can be further divided into academic teaching and Corporate Training, which refers to training in a corporate environment. Applying AR to mobile devices is called Mobile Augmented Reality [5]. This article contains both terms AR and MAR. Both are considered identical and refer to the technology in general. No distinction is made in the article.

Applying AR enhances both learning and teaching and thus, a benefit to education is created. Educational content can be experienced in a new way due to multimodality, interactivity, and engagement. In this context, different learning effects occur. As a learning effect, we understand the change in knowledge, skills, or abilities resulting from learning activities by applying AR. For instance, memory ability, learning motivation, or learning effectiveness can be enhanced [4][6].

However, innovative technologies always relate to new interaction paradigms and, thus, a new experience for the user [7][8]. UX refers to the subjective impression of the user towards a product, system, or service [9]. A positive UX is an essential success factor for interactive products [10]. For this, it is crucial to consider the user's perception of the respective product. Thus, the UX must be measured to provide insights into improving AR and creating a positive UX [11].

In this regard, this article focuses on the UX evaluation of AR. In particular, we analyzed (1) research articles containing an empirical UX evaluation of AR to provide insights into the status quo of UX evaluation. Moreover, we (2) analyzed articles containing a UX model or framework in relation to AR to provide the theoretical foundation. In this article, the authors use *model* and *framework* synonymously.

A Systematic Literature Review (SLR) was conducted to provide these insights into the current state of research in this field. The conducted SLR follows a five-step approach, including five defined scopes based on the Reporting Items for Systematic Reviews and Meta-analysis (PRISMA) guidelines [6][12].

This article is structured as follows: In Section II, we introduce the related work referring to UX and AR. Moreover, the research objective and research questions are explained. In Section III, the methodological approach of the SLR is shown. Results are illustrated in Section IV. Lastly, a conclusion is given in Section V.

## II. Related Work

This Section II introduces the related work regarding UX evaluation. We discuss the concept of UX, its measurement, and common UX questionnaires. Finally, we explain the research objective of this study.

### A. The Concept of User Experience

UX is defined by the ISO as "a person's perceptions and responses that result from the use and/or anticipated use of a product, system or service" [9]. This rather broad description indicates that UX is a multidimensional construct of different dimensions [7][8].

In this context, Usability is strongly related to UX. Usability is defined as the "extent to which a product, system or service can be used by specific users to achieve specific goals with effectiveness, efficiency, and satisfaction in a specific context of use" [9]. Moreover, Usability is declared as a subdimension and, thus, part of the concept of UX.

Different approaches were conducted to break down the construct of UX and get a better understanding. [13] differentiated UX into pragmatic and hedonic properties. Pragmatic aspects refer to task-related, functional factors, whereas hedonic aspects describe emotional factors. This differentiation is a common perspective and many UX researchers rely on this. However, some problems concerning the quantification occur. Pragmatic qualities, such as efficiency, are task-related and, thus, can be measured. In contrast, hedonic factors refer to the emotional perception of the user. In this regard, it is quite difficult to quantify this as there is no specific underlying concept. Moreover, it depends on the specific context whether some quality is pragmatic or hedonic. Hence, it could be difficult to determine and quantify the UX in some cases [8].

Against this background, [8] developed a further distinction. For this, UX was broken down into a set of quality aspects defined as follows: "A UX quality aspect describes the subjective impression of users towards a semantically clearly described aspect of product usage or product design" [8][14].

These aspects can be further applied to quantify and measure UX and, thus, provide insights into the user's perception. This concept is the foundation of several UX metrics. Table I shows the UX quality aspects.

TABLE I. UX QUALITY ASPECTS [8].

| (#) | Factor | Descriptive Question |
|---|---|---|
| (1) | Perspicuity | Is it easy to get familiar with the product and to learn how to use it? |
| (2) | Efficiency | Can users solve their tasks without unnecessary effort? Does the product react fast? |
| (3) | Dependability | Does the user feel in control of the interaction? Does the product react predictably and consistently to user commands? |
| (4) | Usefulness | Does using the product bring advantages to the user? Does using the product save time and effort? |
| (5) | Intuitive use | Can the product be used immediately without any training or help? |
| (6) | Adaptability | Can the product be adapted to personal preferences or personal working styles? |
| (7) | Novelty | Is the design of the product creative? Does it catch the interest of users? |
| (8) | Stimulation | Is it exciting and motivating to use the product? Is it fun to use? |
| (9) | Clarity | Does the user interface of the product look ordered, tidy, and clear? |
| (10) | Quality of Content | Is the information provided by the product always actual and of good quality? |
| (11) | Immersion | Does the user forget time and sink completely into the interaction with the product? |
| (12) | Aesthetics | Does the product look beautiful and appealing? |
| (13) | Identity | Does the product help the user to socialize and to present themselves positively to other people? |
| (14) | Loyalty | Do people stick with the product even if there are alternative products for the same task? |
| (15) | Trust | Do users think that their data is in safe hands and not misused to harm them? |
| (16) | Value | Does the product design look professional and of high quality? |

In the following section, we will introduce common UX evaluation and measurement approaches in UX research.

## B. UX Evaluation and Measurement

Various methods measuring the UX can be found in scientific literature [15][16]. Therefore, the methods cover a wide range of different research objectives and questions. However, the methods differ in terms of the research objective and the application scenario. The following Figure 1 illustrates the most common methods referring to [16]:

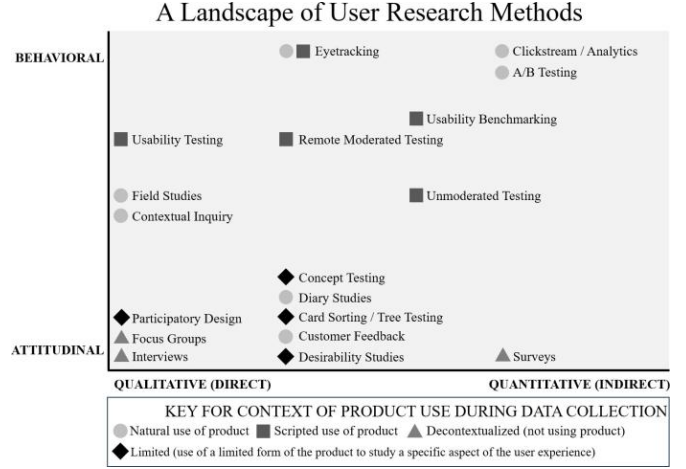

Figure 1. The most common UX Research Methods based on [16].

In general, methods are differentiated into subjective and objective evaluation. Subjective methods relate to self-reported data as direct feedback from the user, e.g., questionnaires. Objective methods refer to analytical data, e.g., eye-tracking data or time measurement. It is common to gather self-reported data as it provides direct user feedback referring to the subjective impression. Applying questionnaires is a simple, fast, and cost-efficient way of collecting self-reported data. Standardized questionnaires are, therefore, the most established method in quantitative UX research.

## C. User Experience Questionnaires

As standardized UX questionnaires are the most common way of collecting self-reported data, we want to introduce them in this section. In general, UX questionnaires aim to gather data on the subjective impression of users. The structure is based on different factors, measurement items, and scales concerning the respective focus. Thus, the construct of UX is broken down by different factors. However, due to the lack of common ground concerning the construct of UX, the UX questionnaires indicate a high heterogeneity. On the one hand, factors can measure the same but are named differently. On the other hand, factors have the same designation but measure something different. Thus, existing UX questionnaires differ on the level of the measurement items and their respective factors [14][17][18].

In the literature, approximately 40 established UX questionnaires can be found [7]. Among them, the User Experience Questionnaire (UEQ) developed by [19] is the most widely used questionnaire [20]. The UEQ was based on the UX foundation in relation to [13] and consists of the following six scales divided into pragmatic and hedonic [19]:

- **Attractiveness**: Overall impression of the product. Do users like or dislike it?

- **Perspicuity**: Is it easy to get familiar with the product and to learn how to use it?
- **Efficiency**: Can users solve their tasks without unnecessary effort? Does it react fast?
- **Dependability**: Does the user feel in control of the interaction? Is it secure and predictable?
- **Stimulation**: Is it exciting and motivating to use the product? Is it fun to use?
- **Novelty**: Is the design of the product creative? Does it catch the interest of users?

The questionnaire was designed to evaluate the holistic impression of interactive products. The different factors consist of semantic differential scales and a 7-point Likert scale [19]. Further information can be found online [21].

However, other questionnaires follow another concept of quantifying the UX. There is a huge variety among the formulation of items. Moreover, no questionnaire can measure all UX factors. Thus, it is important to identify and evaluate the relevant UX factors concerning the respective evaluation object. The User Experience Questionnaire Plus (UEQ+) developed by [22] represents a modular framework that can be individualized regarding the specific evaluation context. Therefore, the UEQ+ consists of 16 UX quality aspects that can be combined to create an individual questionnaire. The UEQ+ is a modular extension of the UEQ and follows the common foundation of UX quality aspects [8] (See Section II-A). Further information can be found online [23].

In the following, we want to specify the research objective of this study.

### D. Research Objective and Research Questions

This article focuses on the UX of AR. The overall research goal is to provide the current state of research concerning UX of AR. More precisely, this SLR follows two directions: (1) We aim to conduct the current state of research regarding UX evaluation. Therefore, we did not specify an application field. Besides this, we aim to collect the respective results of UX evaluation in the field of training and education as this is part of the researcher's doctoral project. (2) we analyzed research articles, including models, frameworks, or reviews in relation to the UX of AR to provide the theoretical foundation of this research topic. Against this background, we address the following research questions:

**RQ1:** *Which methods were applied for measuring UX in the context of AR?*

**RQ2:** *What theoretical models and frameworks exist concerning UX and AR?*

**RQ3:** *What results were conducted in UX research regarding AR in the domain of training and education?*

Based on this, the SLR was conducted. The detailed approach is declared in the following Section III. While speaking from the technology in general within this paper, the term AR also includes the different types such as AR.

### III. METHODOLOGICAL APPROACH

Section III provides an overview of the methodological approach of this SLR. We introduce the specific procedure, scopes, and several stages of the literature search.

### A. Procedure

In the following, we illustrate the methodological approach used in this article. The three authors conducted the research to identify the current state of research concerning the UX of AR. Therefore, a Systematic Literature Review was conducted. The SLR follows a five-step approach (See Figure 2) including five defined scopes (See Figure 3). The processed steps include basic and advanced screening and filtering. Furthermore, a qualitative assessment based on two metrics for selecting articles to ensure the quality of records was applied. Moreover, specific articles in the field of Corporate Training were conducted. The procedure was generally based on the Reporting Items for Systematic Reviews and Meta-analysis (PRISMA) guidelines [12]. We chose this procedure because other approaches for conducting SLR show some limitations, e.g., a lack of explicit guidelines for the quality assessment as well as the insufficient transparency of reporting intermediate results [6]. The detailed approach is shown in the following Figure 2.

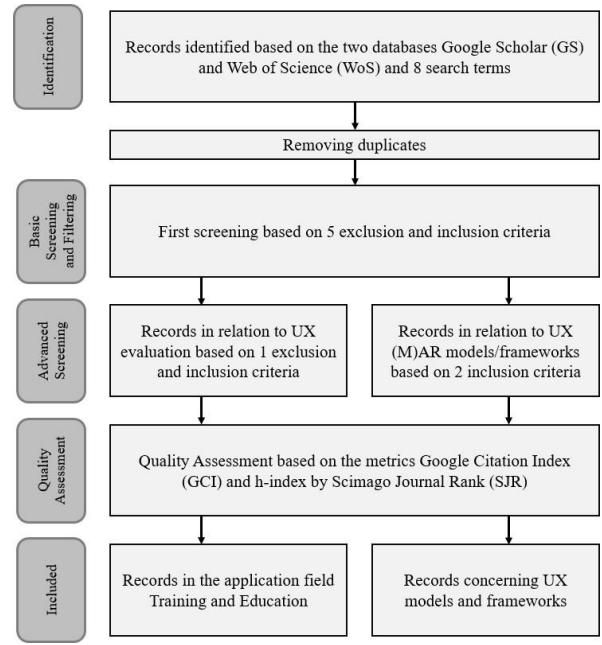

Figure 2. Methodological Approach.

In the following, the different scopes of this study are explained.

### B. Scopes

For this SLR, we have defined five scopes for filtering and screening relevant records regarding the research questions. The five scopes are illustrated in the following Figure 3.

The first three stages aim to identify relevant articles referring to a UX evaluation of AR, including empirically collected data. Stage 4 introduces a specific quality assessment to ensure reliable and valid results within the respective records. Therefore, quality parameters were applied. In the last Stage 5, we particularly analyzed research conducted in the field of Training and Education. The different stages are further described below. The authors processed all the steps.

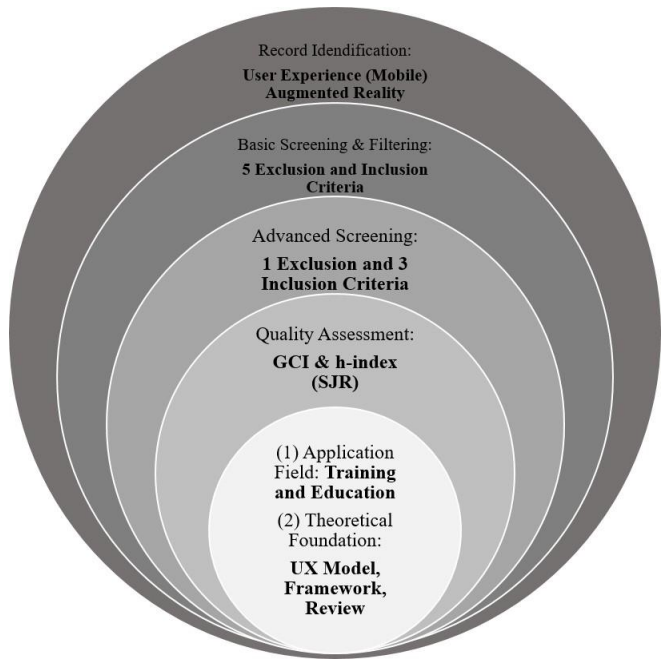

Figure 3. Scopes.

## C. Stages of Literature Search

*1) Stage 1: Identification:* We used **Google Scholar (GS)** and **Web of Science (WoS)** for the literature search and record identification as both are two of the most common and largest databases for scientific research. Thus, both cover mostly all published research articles. We applied eight search terms to both databases. The search terms are composed of the keywords *"User Experience"* and *"Augmented Reality"* and their abbreviations as listed in the following:

- "User Experience Augmented Reality"
- "User Experience Mobile Augmented Reality"
- "UX Augmented Reality"
- "UX Mobile Augmented Reality"
- "UX AR"
- "UX MAR"
- "User Experience AR"
- "User Experience MAR"

We have used the term *"Mobile Augmented Reality"* and its abbreviation to identify all relevant articles. We also included the abbreviations to ensure that all relevant articles were found. In relation to GS, the search terms were applied in the Advanced search performing *"allintitle: first keyword "second keyword"*. Referring to WoS, the keywords were searched using Basic search (Title and Topic). As the focus is on UX evaluation, we assumed UX as the first keyword for the search. Thus, 16 datasets (8 search terms for both databases) were considered. The last searches were conducted on *July 31, 2023*. Based on the 16 datasets, a total number of **498** records was found.

*2) Stage 2: Basic Screening and Filtering:* All duplicates were deleted before basic screening and filtering as the second stage. Afterward, all records were screened based on different inclusion as well as exclusion criteria (see Table II):

For basic screening and filtering, the titles and abstracts of each record were analyzed based on the criteria *in1 - in5*

TABLE II. INCLUSION AND EXCLUSION CRITERIA

| Inclusion criteria | Exclusion criteria for Basic Screening |
|---|---|
| (in1) Focus on UX of AR | (ex1) Focus on VR instead of AR |
| (in2) Accessibility of full-text | (ex2) No accessibility of full-text |
| (in3) Research language English | (ex3) Written in non-English |
| (in4) Peer-reviewed | (ex4) Grey literature |
| (in5) Empirical data collection or theoretical model/framework (also SLR) | (ex5) insufficient information |

and *ex1 - ex5* described in Table II. We specifically included all papers focusing on the UX of AR. We analyzed whether the full text was available or not. We only considered English language literature as the official research language. Lastly, we only included peer-reviewed records. Grey literature, such as white papers, theses, etc., were excluded. This results in a total of **223** records.

*3) Stage 3: Advanced Screening:* In the third stage, an advanced screening was conducted applying the criteria *in6, ex6, in7, in8* (See Table III). Therefore, the criteria *in6* and *ex6* refer to records concerning the empirical evaluation, whereas *in7* and *in8* relate to the records for the theoretical foundation.

TABLE III. INCLUSION AND EXCLUSION CRITERIA FOR ADVANCED SCREENING

| Inclusion criteria | Exclusion criteria |
|---|---|
| (in6) UX/Usability evaluation goal | (ex6) Lack of focus in UX/Usability evaluation goal |
| (in7) UX model/framework included | |
| (in8) Systematic Literature Review | |

In particular, the abstracts and full texts of the records were analyzed to determine whether the primary goal of the respective study in relation to the research questions was addressing UX/Usability. Moreover, we considered whether an empirical study collecting empirical data was conducted. A number of **121** can be provided.

Furthermore, we analyzed whether a UX model or framework was proposed or contained regarding the identification of the theoretical foundation. As a result, **12** articles were identified.

*4) Stage 4: Quality Assessment:* In stage four, we employed a qualitative assessment to identify articles with high impact in the research field. Therefore, we applied two measures - the Google Citation Index (GCI) and the h-index provided by Scimago Journal Rankings (SJR). Even though both metrics show certain limitations, they are increasingly applied for the indication of paper impact [6][24][25]. Based on these metrics, we aimed to provide qualitative results.

Firstly, all records regarding UX evaluation were processed. We made a record classification according to their type *(B = book chapter, J = Journal article, C = conference proceedings, A = none of the three)*. In the following, the types and the respective record number is shown:

- **B** = Book chapter: **6** records
- **J** = Journal article: **54** records
- **C** = Conference proceedings: **61** records

In the second step, we conducted both metrics for all records. For the h-index, we looked up the specific scores of the respective publisher concerning each article. If the h-index of the latest published issue was missing, we used the closed available score of the previous issues. Afterward, we calculated the median h-index of each type. We used the median value rather than the mean due to large ranges of the respective scores [6]. Thus, the median h-index represents the threshold for quality classification.

Concerning the GCI, we computed the *Average Citation Count (ACC)* of each record by dividing the overall citation count by the number of years [6][26]. We also calculated the median score of overall citations as the relevant threshold. Thus, all records with an ACC above the median were considered qualitative. As all records with the type **A** do not have any citations and the h-index does not apply to them, these articles were excluded. The different median scores are shown in the following Table IV:

TABLE IV. MEDIAN VALUES OF RECORDS REGARDING UX EVALUATION

| Median | B | J | C |
|---|---|---|---|
| GCI | 1.62 | 2.5 | 1.5 |
| h-index | 0 | 46 | 7 |

Both metrics with the respective thresholds were applied, resulting in a final batch of **71** records. The resulting distribution by type is divided as follows:
- **B** = Book chapter: **2** records
- **J** = Journal article: **31** records
- **C** = Conference proceedings: **38** records

All records are listed in the appendix V.

In relation to theoretical models and frameworks, 12 records were conducted with the following distribution:
- **B** = Book chapter: **1** records
- **J** = Journal article: **3** records
- **C** = Conference proceedings: **7** records
- **A** = other type: **1** records

Due to the large discrepancy in the metrics values, the quality assessment for the eleven articles was rather difficult. However, all articles have at least five citations. Thus, the eleven articles were further analyzed.

*5) Stage 5: Included records:* In the final stage 5, we made a further record selection and examined all articles in relation to the application field of *Training and Education*. This was done due to the specific research focus of the authors. This results in a number of **18** records. For this, we considered the respective UX evaluation results of these articles.

Furthermore, all records concerning a theoretical foundation were analyzed. This results in a final record batch of **30**. Figure 4 shows the approach with the respective number of records in each stage.

In the following, the results are illustrated.

IV. RESULTS

This section presents the results of this SLR. The section IV-A gives an overview of general information concerning research. We examined the records based on the pattern by (1) publication year, (2) origin, (3) application field, (4) application device/hardware, (5) software, and (6) methodological approach. For (1) and (2), both records concerning UX evaluation (*n = 71*) and UX models/frameworks (*n = 12*) identified in Stage III-C4 (Stage 4) Quality Assessment were examined. For (3) - (6), only the 71 UX evaluation articles were considered.

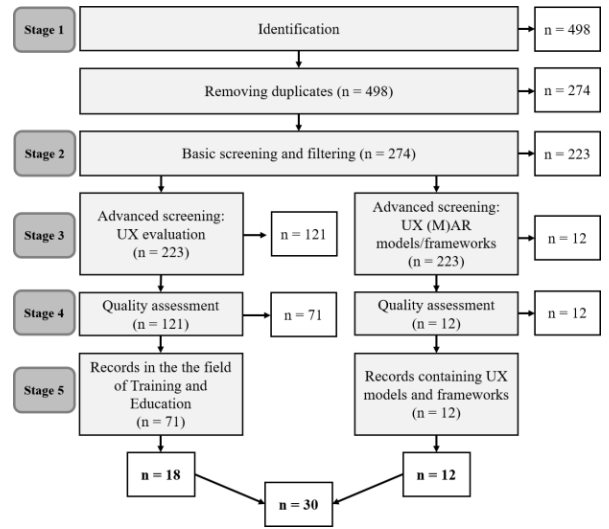

Figure 4. Methodological Approach.

To provide a deeper insight into the UX of AR in the field of Training and Education, as well as the theoretical foundation, the articles resulting from Stage III-C5 (Stage 5) were further considered. Section IV-B illustrates the respective UX evaluation results in the field of *Training and Education*. Therefore, the *18* full articles were analyzed. Lastly, Section IV-C shows all records, including a UX model, framework, or review, to provide an overview of the theoretical foundation concerning UX of AR. The respective *12* were analyzed for this.

*A. General Information*

*1) Pattern by Year:* The earliest publication year records (n = 5) were identified is 2011. Until 2017, record numbers were rather low. Since 2018, there has been a strong increase in articles. The highest number of publications reported in 2019 was 12. Figure 5 illustrates the number of records over the years.

*2) Pattern by Origin:* Regarding the origin, the identified records are spread across 4 continents. Most studies (n = 40) were conducted by researchers from Europe, followed by Asia with a total of 24. One article can be classified as Asia and Europe. Two records are located in North America, whereas four articles are assigned to South America. No articles could be identified in relation to Africa or Australia. Figure 6 presents the paper distribution by origin.

The 71 identified records regarding the UX evaluation of AR are considered in the following. It must be noted that some articles conducted several studies within one research article. Therefore, it is possible that the numbers of the respective results do not correspond to the numbers of the identified papers.

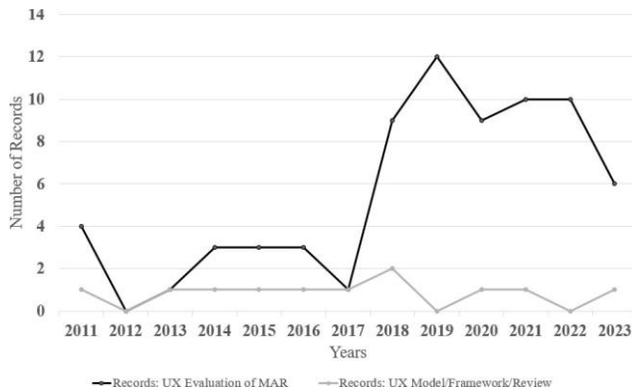

Figure 5. Number of Records Regarding the Year.

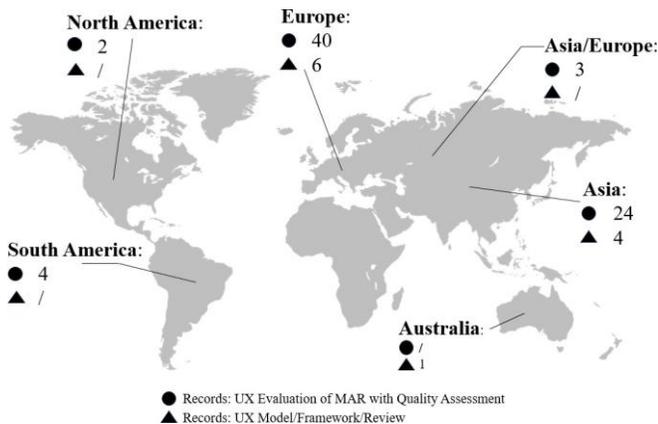

Figure 6. Records by Origin.

*3) Pattern by Application Field:* We categorized the records according to the specific application domain. Therefore, the researchers examined the papers and clustered the different domains into six main classes:

- Training and Education (n = 18)
- Marketing / Commercial Applications (n = 16)
- Culture Heritage / Museum (n = 18)
- Entertainment (n = 5)
- Medicine (n = 3)
- Navigation (n = 11)

Most records were collected in the field of training and education. As already described, the field can be further divided into academic teaching and Corporate Training. Therefore, ten records can be identified regarding academic teaching, whereas eight articles refer to Corporate Training scenarios.

The further major domains are Culture Heritage / Museum, with a total 18, and Marketing / Commerical Applications, with 16 identified articles. Additionally, applying AR for Navigation could be recorded eleven times. Lastly, five records in the field of entertainment and three in medicine were conducted.

*4) Pattern by Application Device:* Concerning AR, the technology can be applied to different hardware settings. By observing the literature, Mobile Augmented Reality often refers to Mobile Devices [5][6]. However, this is not always clear. For instance, AR glasses are also mobile in the sense of their property as they are not firmly linked to a place/position. Thus, we made a distinction between *handheld* and *head-mounted* in this study.

Results indicate that most studies have used handheld devices. This was determined for 48 articles. Android was used 28 times, and iOS 10 times as an operation system.

In comparison, 14 records applied head-mounted displays. Against this, the Microsoft HoloLens was used ten times, whereas the MagicLeap was implemented four times.

One study used a Web AR application. Eight studies did not specify the application device.

In the end, we took the respective underlying development software into account. This resulted in a very opaque overview, which was not specified for 43 articles. However, it can be noted that Unity is the most commonly used platform for prototype development, extended with different common AR development PlugIns, e.g., Vuforia, or its own programming.

*5) Pattern by Methodological Approach:* As described in Section II-B, various methods for UX evaluation can be found. The respective methods have different characteristics concerning their approach. Lastly, we examined the records' methodological approach to provide insights into the evaluation method. Results show that 38 articles applied purely quantitative methods. In contrast, ten records used a qualitative method. A total of 23 conducted a mixed-method approach applying both quantitative and qualitative methods. Questionnaires (n = 69), interviews (n = 19), (Usability) performance analysis (n = 2), observation logs (n = 2), NLP approaches (n = 2), Thinkin'-Aloud (n = 1), and eye-tracking (n = 1) can be listed as applied methods. For this, questionnaires are the most commonly used method.

As quantitative UX evaluation in questionnaires is the most established, we further examined the existing applied questionnaires. In summary, 40 records include individualized developed questionnaires. In contrast, 29 studies applied standardized, existing questionnaires. Regarding existing ones, eleven metrics could be identified. These include the UEQ [19], SUS [27], QUIS [28], AttrakDiff [29], SSQ [30], NASA TLX [31], TPI [32], HARUS [33][34], PSSUQ [35], TAM [36], and UTAUT [37], although the latter two originally belong to the field of technology acceptance research. Most of them were only used one to three times. Only the UEQ and the SUS have been applied more frequently. The SUS was applied eight times, whereas the UEQ was applied ten times. Therefore, it can be stated that the UEQ is the most widely used UX questionnaire.

Up to here, we considered relevant records resulting from stage four (See Section III-C4) to provide a comprehensive overview into general insights concerning the UX evaluation of AR. In the next Section IV-B, we took papers resulting from Stage 5 (See Section III-C5 into account to present details about UX evaluation in the field of Training and Education as well as the theoretical foundation.

*B. User Experience Evaluation Results in Training and Education*

In this section, we present results regarding the UX evaluation of AR in the field of Training and Education. In particular, **18** records were identified referring to Training and Education. As shown above (See Section IV-A3), the articles can be further classified into academic teaching (n = 10) and

Corporate Training (n = 8). The relevant articles are shown in the Appendix (See Table VI). In the following, we provide detailed results in relation to these papers.

*1) Records in the Field of Corporate Training:* For the field of Corporate Training, the records [38]–[45] were identified. [38], [42], and [39] conducted a quantitative method whereas [40][41][43]–[45] applied a mixed-method approach. In particular, Questionnaires (n =10), qualitative Interviews (n = 2), and Observation (n = 1) were applied as research methods. Individualized and standardized questionnaires were applied to the questionnaires. For standardized UX questionnaires, the (1) SUS [27], (2) QUIS [28], (3) NASA-TLX [31], (4) AttrakDiff [13][29], and (5) UEQ [19] were identified. Besides these, the Smart Glasses User Satisfaction (SGUS) questionnaire [46] was implemented.

Okimoto et al. conducted a study about welding simulation training with a group of twelve students. Participants perceived the applications as motivating for training due to the novelty of the technology. Introducing and using the application was reported as satisfactory. However, difficulties regarding the visual accommodation were mentioned. In particular, the initial sense of spatial depth and perception of the manual movement can be stated [38].

Helin et al. evaluated an AR system for astronauts' manual work. Results indicate that the use of AR in daily operations is usable. The system was working properly without delays. Based on the quantitative evaluation, the application was considered nearly acceptable. Furthermore, participants could accomplish their tasks quite well. It was perceived as easy to learn new tasks. The attention was captivated positively. However, the experience was perceived as not natural. Moreover, it was not easy to follow the instructions. Information seemed to be hard to read, confusing, and inconsistent. The overall rating was quite positive. Suggestions for improving the UX mainly refer to detailed information and feedback regarding the system control. Besides this, the participants indicated concern in relation to the quality and accuracy of the displayed content [38].

Alenljung and Lindblom evaluated a prototype for assembly instructions with a sample of five participants. The authors set up nine UX goals. The procedure was conducted in two rounds. Results indicate that the majority of the UX goals were not met. The problems mentioned here are not properly working AR function, the clarity of instructions, the sharpness in content projection, and the object detection [45].

Aromaa et al. applied AR in maintenance processes. A positive UX was evaluated and the system was well accepted. It was perceived as useful and supportive. Moreover, it was rated as flexible, effective, wonderful, and satisfying. The system's visual appearance was liked and easy to use [40].

Both studies by Thomaschewski et al. evaluated an AR system supporting the temporal coordination of spatially dispersed teams. 22 participants took part in this study. In this context, Usability and UX evaluation were applied to identify the best interface. Thus, no further derivations can be reported [43][44].

Heo et al. provides insights into the UX evaluation of an AR-based automobile maintenance content application using a mobile device and the HoloLens. The study was conducted on 44 participants. In particular, the Awareness, Comfort, Functionality, and Space Perception were conducted and compared. Results show that the UX was not significantly different between the mobile device and the AR glass. Comfort was experienced more positively on the mobile device. In contrast, space perception of the AR glass showed better results [42].

Scavo et al. explored AR for telementoring. Twelve participants took part in the study. The application was perceived as greatly stimulating and playful, improving engagement. Results indicate that the interaction was intuitive, natural, and fast to learn [39].

To sum up, the results of the study report rather positive UX evaluation results concerning pragmatic as well as hedonic aspects. Thus, AR can be a benefit in a Corporate Training environment. However, it can be shown that problems often occur with the system functions of the AR applications.

*2) Records in the Field of Academic Teaching:* For the field of Academic Teaching, the records [47]–[56] were conducted. [50][51][53][55][57] and [52] conducted a quantitative approach. In comparison, [47]–[49] and [56] applied a mixed-method approach. Only [54] followed a qualitative approach by applying NLP techniques to analyze app reviews and ratings. Similar to the field of Corporate Training, Questionnaires (n = 10), qualitative Interviews (n = 3), and Observations (n = 1) were conducted. Applied questionnaires are individualized as well as standardized. Among the standardized UX questionnaires, the (1) UEQ [19], (2) PSSUQ [35], and (3) NASA-TLX [31] were used. Besides these, the Technology Acceptance Model (TAM) [36], the Emotions Questionnaire, the Temple Presence Inventory [32], and the Simulator Sickness Questionnaire [30] were used for evaluation.

Redondo et al. conducted a case study applying AR for spatial analysis in urban design processes. A student satisfaction survey and the Students ' Evaluation of Educational (SEEQ) questionnaire were applied for evaluation. Results indicate that AR can improve performance, shorten project development time, and promote creativity. However, hardware restrictions in object registration and high implementation costs are concluded [48].

Pribeanu and Iordache evaluated the usability of an AR-based learning scenario focusing on motivational value in a chemistry learning environment. Results show that AR application is perceived as supportive, exciting, motivating, and easy to use. The visualization and user guidance were perceived as positive. As negative aspects, the representation and augmentation of the educational content were difficult to distinguish [47].

Sarkar and Pillai considered user expectations toward AR. The authors developed expectations based on the three dimensions (1) content, (2) incentive, and (3) interaction in relation to learning with AR. Relevant characteristics concerning the expectations were defined for each of the three dimensions. These include Visual Cues, Informative, Situational Regeneration, and Dynamic for (1), Developing Interest, Cognitive Sustenance, Creative Instances, and Playfulness for (2), and Immersive, Tangible, Familiarity, and Exploratory for (3) [49].

Kazanidis and Pellas conducted a study applying AR in Mathematics. In summary, participants perceived the learning procedure using AR as positive. The AR application was perceived as visually appealing, helpful, and easy to use. Moreover, participants considered it to be exciting and useful. As negative aspects, a longer period of familiarization concerning AR, less effectiveness, and marker recognition, which did not work, was reported [50].

Kim-Berman et al. developed and tested a virtual tooth identification test. The evaluation shows problems in viewing and manipulating the AR application. Moreover, a high loading time and battery consumption could be determined. Nevertheless, the application was evaluated as effective [51].

Smaragdina et al. studied the UX of computer graphics educational comics applying markerless augmented reality. Evaluation results show a positive UX among all six scales of the UEQ. Pragmatic qualities were rated higher than hedonic properties [52].

Vrellis et al. applied UX and technology acceptance measures to evaluate an AR application for science literacy. Results indicate moderate spatial presence, low simulator sickness, and high acceptance as well as satisfaction [53].

Dominguez Alfaro and Puyvelde investigated the UX of AR apps by analyzing app reviews. Results show that technical issues, features, and user instructions must be improved [54].

Liu et al. evaluated a web-based AR learning tool. Study results indicate that a positive attitude toward the technology enhances the experience. Furthermore, the factors Perceived ease of use, Attitudes toward technology use, Need for technological pedagogical content knowledge, Experience with technology–Traditional, Behavioral intention–Traditional, and Behavioral intention–Innovative show a positive significant effect on the UX [55].

Lastly, Santana et al. took the UX of a learning app into account, investigating the overall satisfaction, technological acceptance, mental workload, and emotional response. Therefore, the UTAUT and the PSSUQ were applied. Observations show that the application improves learning quality. Moreover, high engagement and satisfaction could be determined [56].

In summary, AR applications in academic teaching indicate positive UX evaluation results. Therefore, both pragmatic as well as hedonic qualities are perceived as positive. Overall, learning and teaching activities can be enhanced. In general, deficiencies and errors in the system features and functionalities of the applications were cited as negative. However, a high potential is cited among all records.

*C. Theoretical Foundation concerning UX of AR*

The second research objective was the presentation of the theoretical foundation. We examined records, including models and frameworks concerning UX of AR. We further included relevant reviews or studies dealing with the theoretical foundation to provide a comprehensive overview. In summary, *12* records could be identified. *6* articles contain specific UX models and frameworks whereas *6* records refer to reviews or papers conducting general insights. The results are presented in the following section.

*1) UX Models and Frameworks:* In summary, [58]–[63] contain a UX model or framework.

Ritsos et al. developed a theoretical evaluation framework aiming to measure important aspects of the development of AR applications in a standardized way. The framework was classified into different categories input (visual, auditory, tactile, and kinæsthetic), output (visual, auditory, and haptic), context awareness, sense of immersion, health, safety and integrity, and privacy and security [58]. The framework is shown in the following Figure 7:

Irshad and Rambli presented an early framework focusing on the design and evaluation of the UX of AR in 2015

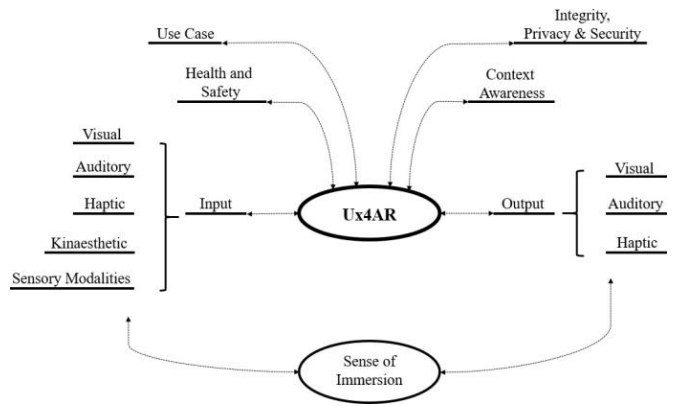

Figure 7. UX4AR Theoretical Framework Mapping by [58].

[59]. Based on this, the authors developed the multilayered conceptual framework for the enhancement of the UX of AR. The proposed framework put together the relevant factors for designing AR applications. The framework introduces the four layers *AR Product/Service Features*, *Time*, *Specific Context*, and *UX* [60]. The framework is illustrated in Figure 8

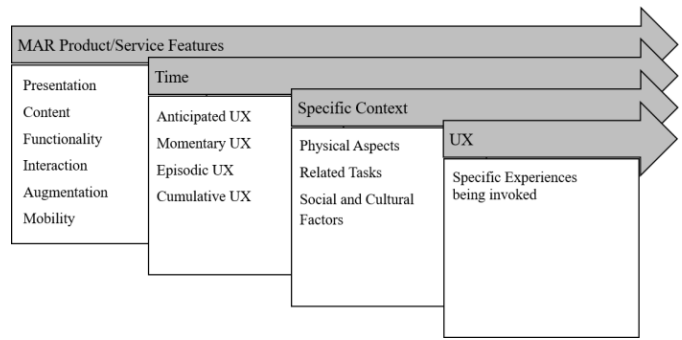

Figure 8. Multilayered conceptual framework for enhancing the UX of MAR products by [60].

Irshad et al. further proposed a UX Design and Evaluation framework concerning MAR. The evaluation differs between instrumental and non-instrumental quality attributes. Furthermore, the specific domain of the AR application is considered [64]. Figure 9 shows the evaluation framework.

Han et al. developed a UX model in the context of urban heritage tourism [65]. The authors identified relevant factors within this application field and extended the UX model by [13].

Lastly, Ahmad et al. developed a preliminary model focusing on the emotional UX applying the *Kansei* engineering approach. The so-called Augmented Reality Mobile Application Design (ARMAD) model consists of the three components *User Interface Design*, *Interaction Design*, and *Content Design*. The aim was to achieve a pleasurable design for the user [63]. The model is presented in the following Figure 10:

Besides this, Irshad et al. reviewed MAR studies from a UX perspective [5]. As a result, the authors identified the four relevant records [13][66]–[68]. However, a distinction must be made between these papers. The models by Hassenzahl and Ja¨a¨sko¨ and T. Mattelma¨ki were generally developed without reference to AR [13][67]. For example, the framework developed by [13] is one of the most popular ones in UX research and, e.g., the base for the quantification of UX in some metrics. Hence, both UX frameworks can be applied to different contexts. In contrast, the frameworks developed by [66] and [68] refer specifically to MAR. Thus, we only discuss the latter two.

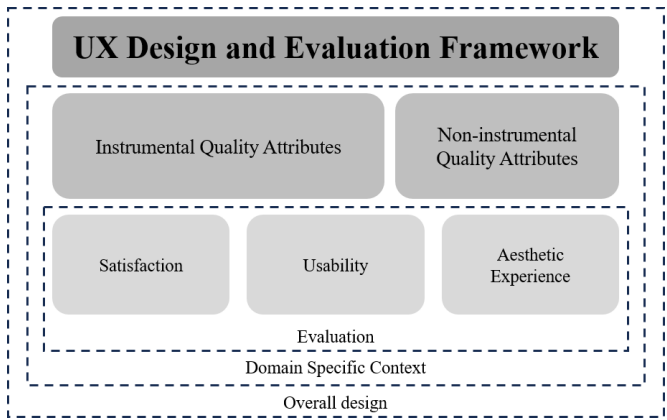

Figure 9. UX Design and Evaluation Framework for MAR by [64].

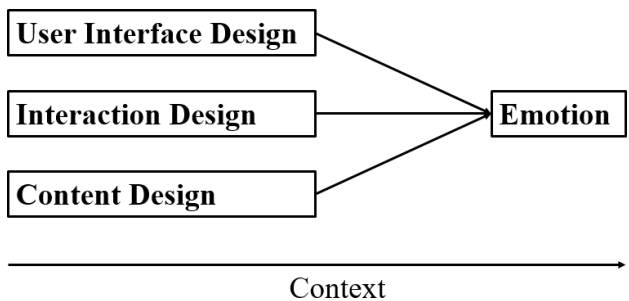

Figure 10. Augmented Reality Mobile Application Design (ARMAD) Model by [63].

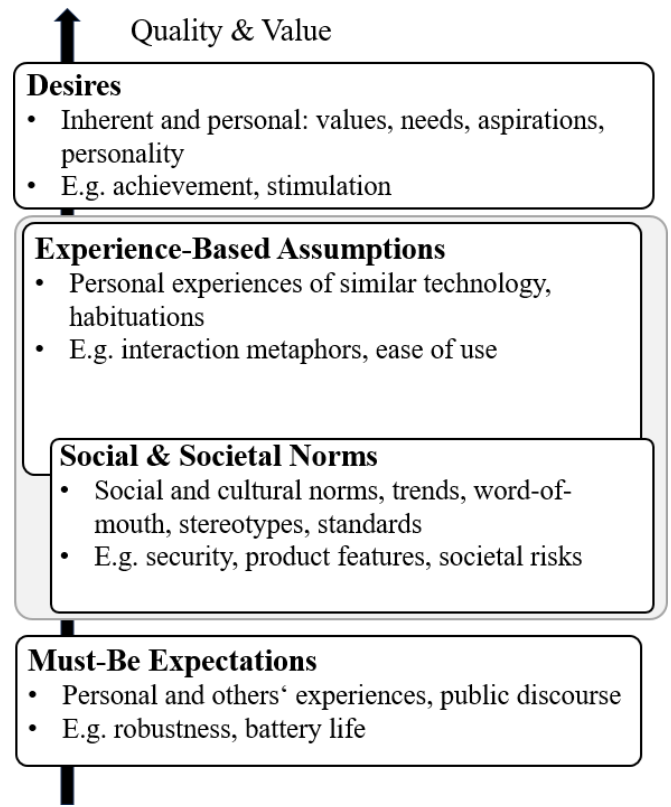

Figure 11. Early framework by [66].

Perritaz et al. developed a framework focusing on the deployment of real-time concerning AR. In particular, the authors introduced the factor *Quality of Experience* consisting of the variables real-time adaption, frame rate, and image size. Their proposed framework aims to improve the *Quality of Experience* and, thus, the UX by measuring these variables and improving the collaboration in real time. This can be considered a more mathematical approach [68].

Furthermore, Olsson presents an early framework based on different layers of user expectations that people have with new technologies such as MAR [66]. The work-in-progress framework was based on the four layers *Must-Be Expectations*, *Social and Societal Norms*, *Experience-Based Assumptions*, and *Desires*. The layers describe the origin of the different user expectations towards technology. Figure 11 illustrates the early framework.

As a result, only a limited number of models can be presented. The models have different characteristics in terms of focus. Concerning a common base, the UX model by [13] was applied and extended. Moreover, no model provides a combination of relevant system characteristics of AR and UX factors based on relevant literature concerning training and education. The following discusses the **6** relevant review records.

*2) Review Paper:* To sum up, [6][69]–[73] provide insights into UX of AR as a review paper.

Arifin et al. examined existing research concerning UX metrics for AR applications. Additionally, the authors specifically considered the field of Education as there are no standard measurements. Results showed that there is no metric measuring the UX from AR applications in the field of Education [69].

Irshad and Ramdbli conducted an initial study concerning the UX of MAR. The authors performed a further review presenting a comprehensive overview of the advances of MAR from a UX perspective [72]. Therefore, nine UX studies of AR and four studies in relation to a UX framework for MAR were identified (See Section IV-C1). The authors identified a lack of user research in this field [70].

Ghazwani and Smith examined the AR interaction from a UX perspective. It is argued that three aspects are relevant: (1) user type, (2) user interface, and (3) virtual content [71].

Law and Heintz conducted an SLR on the topic of AR application in education from a usability and UX perspective. Results show a lack of innovative AR-specific usability/UX evaluation methods. We will refer to this paper again in a follow-up chapter (See Section IV-B) [6].

Lastly, we want to present [73] summarizing the results of past studies in this article. The article presents an overview of User Expectations that are relevant for MAR. In summary, 16 User Expectations (categories) were identified in previous studies [74]. The Expectations are classified into the six classes

*(1) Instrumental Experiences (IE)*, *(2) Cognitive and Epistemic Experiences (CEE)*, *(3) Emotional Experiences (EE)*, *(4) Sensory Experiences (SeE)*, *(5) Social Experiences (SoE)*, and *(6) Motivational and Behavioral Experiences (MBE)*. The classes and categories regarding MAR were identified in The 16 categories are listed below [73]:

- (1) Empowerment (IE)
- (2) Efficiency (IE)
- (3) Meaningfulness (IE)
- (4) Awareness (CEE)
- (5) Intuitiveness (CEE)
- (6) Amazement (EE)
- (7) Surprise (EE)
- (8) Playfulness (EE)
- (9) Liveliness (EE)
- (10) Captivation (SeE)
- (11) Tangibility and Transparency (SeE)
- (12) Collectivity and Connectedness (SoE)
- (13) Privacy (SoE)
- (14) Inspiration (MBE)
- (15) Motivation (MBE)
- (16) Creativity (MBE)

The author further provides practical instruments for the evaluation. Therefore, summative and formative measures concerning different experience categories were presented. The summative statements refer to the overall evaluation of the MAR application and are more general. In comparison, the formative statements are detailed and focus on the quality of specific features and their direct influence on the UX. For further details concerning the statements, we refer to [73].

### D. User Experience Augmented Reality Questionnaire

As shown, quantitative UX measurement is the most widespread. In particular, the application of standardized questionnaires is a common way to measure the UX as it is fast, simple, and cost-efficient. Different standardized UX questionnaires can be found in scientific literature [7]. However, the questionnaires have different characteristics, structure, and focus. Thus, not every questionnaire is applicable to AR in CT. Concerning literature, three standardized UX questionnaires evaluating AR can be found. The three questionnaires are listed in the following:

- (1) **Handheld Augmented Reality Usability Scale (HARUS)** [34][33]
- (2) **Augmented Reality Immersion (ARI) Questionnaire** [75]
- (3) **Customizable Interactions Questionnaire (CIQ)** [76]

These three AR-specific questionnaires already clearly show the differences in structure and focus. The Handheld Augmented Reality Usability Scale (HARUS) developed by [33][34] focuses on the usability evaluation of handheld AR devices. To be more precise, it considers the ergonomic and perceptual issues of handheld AR issues. The structure is based on the two factors *manipulability* and *comprehensibility*. *Manipulability* refers to the ease of handling the device, whereas *comprehensibility* refers to the ease of understanding the presented information. Each factor comprises eight items based on a seven-point Likert scale (1 is "strongly disagree" and 7 is "strongly agree") [33][34].

The Augmented Reality Immersion Questionnaire (ARI) was developed by [75] focusing on immersion in location-aware AR settings. Immersion is described as cognitive and emotional absorption during the use of interactive media such as AR. The questionnaire consists of the three factors *Engagement*, *Engrossment*, and *Total immersion*. These factors are further broken down into six sub-factors: *Interest*, *Usability*, *Emotional attachment*, *Focus of attention*, *Presence*, and *Flow* with a total of 21 items based on a seven-point Likert scale [75].

The Customizable Interactions Questionnaire (CIQ) measures the subjective impression focusing on the quality of interaction with objects in AR scenarios. The questionnaire developed by [76] contains five factors *Quality of Interactions*, *Comfort*, *Assessment of Task Performance*, *Consistency with Expectation*, and *Quality of the Sensory Enhancements* with a total of 17 items based on a five-point Likert scale [76].

Based on this, the heterogeneity of the different questionnaires emerges. Up to now, no other domain-specific standardized UX questionnaire regarding AR can be found. Moreover, no questionnaire for the Training and Education scenarios could be found.

### E. User Experience and Learning Effect

We finally want to present the state of research referring to UX evaluation and learning effect in relation to AR in Training and Education. However, only a little research on this topic can be found [6][77].

Law and Heintz conducted an SLR on the usability and UX of AR applications for K12 education. Within this research, the authors also examined records referring to the relationship between UX and the learning effect. Results show that only a few studies describe the relationship between UX and learning effect qualitatively. Only Lin et al. computed the correlation between both factors [78]. For this, low, insignificant correlations could be identified [6].

Knowledge/skill-specific tests and systematic observations were applied as common methods for learning effect measurement.

Additionally, Subandi et al. can be stated concerning this topic. Within this study, MAR was applied as a learning solution for vocational students on a topic related to computer network devices. Results show that MAR improves learning effectiveness while implementing a positively evaluated prototype referring to the UX. However, the UX and learning effectiveness were considered separately [77].

No model/framework addressing the relationship between UX and learning effects could be identified. A lack of research can be stated.

## V. DISCUSSION AND CONCLUSION

This article presents a Systematic Literature Review presenting detailed insights into the UX of AR. In particular, we aimed to provide results concerning the UX evaluation of AR in general, especially in the field of training and education, as well as the theoretical foundation. The SLR was based on a five-step approach with five defined scopes. The procedure was derived from the PRISMA guidelines [6][12]. In summary, 498 records referring to eight search terms applied in the two databases, Google Scholar and Web of Science, were identified. As a result, 71 articles regarding the UX evaluation of AR were analyzed after four exclusion stages. In the fifth stage, we identified 18 records referring to the field of Training and Education and 11 articles regarding the theoretical foundation. The main implications of this study concerning the research questions are discussed in the following.

*A. Implications*

In this Section, we discuss the proposed research questions (See Section II-D) by referring to the results (See Section IV of this SLR. Based on the results, we want to derive practical implications for the UX evaluation of AR. Moreover, we address topics related to UX evaluation.

*1) Research Question: Which methods were applied for measuring UX in the context of AR?:* No established method measuring the UX for AR in the field of Training and Education could be identified. Only three AR-specific standardized UX questionnaires could be identified. However, none of the questionnaires specifically focuses on the field of Training and Education. They are rather heterogeneous, focusing on different subsets of the UX. Thus, there is a lack of research concerning UX measurement approaches for AR in this application field.

In general, the most widely used methods are quantitative. More precisely, questionnaires are most commonly applied. For this, both standardized and individualized UX questionnaires were used. It is common to apply more than one questionnaire to expand the study by examining several UX subsets. However, a purely quantitative approach also entails limitations in relation to the evaluation results. When using questionnaires, purely numerical evaluation results are provided. This results in limitations in the interpretability of the results for the researchers. Thus, it is common to conduct a mixed-method approach. Quantitative approaches are often extended by qualitative methods, e.g., interviews. This is being done to obtain the most comprehensive results regarding user perception and to overcome the problem with purely numerical data.

In this regard, we want to raise another line of thought. The various quantitative and qualitative methods are, of course, valid on their own. However, there are no findings as to which specific methods work best together. Furthermore, it is also unclear how the results can be interpreted together and how they fit together. Moreover, the evaluation results are data from the user's perspective. This illustrates that the results require further interpretation by the product designers and developers to derive useful improvement suggestions. Up to now, there has been no approach addressing these points. Nevertheless, this would be an interesting idea to be considered. An approach based on combining quantitative and qualitative results might be assessed. In addition, a bridge could be created between the user's perspective and the developer's perspective. This would enable the developer to derive specific improvement options more quickly and easily based on the results from the user's perspective.

**Practical Implications regarding RQ1:** As UX is a multidimensional construct, it is important to determine the respective evaluation objective. Based on this, the evaluation method must be selected. As shown in the related work (See Section II), it is common to apply standardized questionnaires. Due to the heterogeneity of both holistic and AR-specific existing questionnaires, it is essential to choose the best-fitting questionnaire concerning the evaluation objective. Against this background, customizable evaluation frameworks, such as the UEQ+ [23], could be helpful as researchers can select the relevant scales. Moreover, results should be interpretable and categorizable within the field. In other words, the result should be in the form of a UX score that allows a conclusion to be drawn about users' perceptions. For this, a questionnaire benchmark is useful [79][80]. To gather deeper insights, applying further qualitative methods is common and useful.

*2) Research Question: What theoretical models and frameworks exist concerning UX and AR?:* To sum up, six theoretical UX models/frameworks regarding the UX of AR could be identified. It must be noted that the different approaches refer to different focuses. Moreover, none of the frameworks refer to the same foundations. However, a common line within the research can be shown at [66][73][74] and [5][59][60][64]. Olsson refers to one of the most established theoretical foundations concerning a general perspective toward UX developed by [13][66]. Research provided by [64] is partly based on [66]. Thus, the UX model by Hassenzahl is most commonly used as a foundation [13]. However, no model exists combining relevant AR system characteristics and UX factors based on relevant literature regarding training and education.

**Practical Implications regarding RQ2:** The models can serve as an explanatory concept. Moreover, such a model can be useful in UX measurement [81]. However, no established model exists addressing AR and UX in combination. Research on both UX in general and UX of AR refers to the UX model by [13]. Established UX measurement approaches, e.g., the UEQ [19][21], are based on this model. Thus, this model can be considered as common ground towards UX theory.

*3) Research Question: What results were conducted in UX research regarding AR in the domain of training and education?:* Consistent results regarding the UX evaluation of AR can be seen in the field of training and education. The UX evaluation is often individualized, and methods are applied with a specific focus regarding the respective research objective. The UX is perceived consistently as positive. Both pragmatic and hedonic UX quality aspects are evaluated as highly positive. Negative evaluation results refer to errors and deficiencies concerning the system or functionality of AR. Among all records, applying AR in training and education is indicated to have great potential.

**Practical Implications regarding RQ3:** The main issue is the functionality of the AR system, which was also investigated in previous research [82]. Errors and deficiencies are perceived as negative. One of the reasons for this is that, in most cases, users are using AR for the first time. Thus, practitioners should focus on an error-free running system.

*4) UX Evaluation and Learning Effect:* In recent years, research has been conducted to analyze the learning effects of using AR in Training and Education. As already shown, a positive UX is an essential success factor for interactive products such as AR. Thus, there could be a relation between the UX and the resulting learning effects. However, only a little research on this topic can be found [6][77].

Law and Heintz conducted an SLR on the usability and UX of AR applications for K12 education [6]. Within this research, the authors also examined records referring to the relationship between UX and the learning effect. Results show that only a few studies describe the relationship between UX and learning effect qualitatively. Only Lin et al. computed the correlation between both factors. For this, low, insignificant correlations could be identified [78].

Knowledge/skill-specific tests and systematic observations were applied as common methods for learning effect measurement.

Additionally, Subandi et al. can be stated concerning this topic. Within this study, MAR was applied as a learning solution for vocational students on a topic related to computer network devices. Results show that MAR improves learning effectiveness while implementing a positively evaluated prototype referring to the UX. However, the UX and learning effectiveness were considered separately [77].

No model/framework addressing the relationship between UX and learning effects could be identified, indicating a lack of research.

To sum up, research about the quantification of the potential of applying AR in Training and Education is rather rudimentary. Only a little research was done examining the correlation between a positive UX and learning effects resulting from the application of AR. Lin et al. are the only study that computed the correlation between UX and learning effects. As a result, a low, insignificant effect was conducted [78]. Thus, it has not yet been researched whether a positive UX of an AR application is related to learning effects. A lack of research can be noted.

*5) UX Evaluation and Generative Artificial Intelligence:* Due to its current relevance, we would also like to address the topic of Generative Artificial Intelligence (GenAI). The rapid development of Large Language Models (LLMs), e.g. ChatGPT [83], impacts various research fields, including UX research. Based on their structure, LLMs show a strong ability to understand and generate natural language. Such models are useful in deep learning and natural language processing tasks. Thus, LLMs enable new opportunities to enhance, support, and automate activities along the research process. Results from the latest research show a considerable potential for applying GenAI in UX research [18][84][85]. Future research, therefore, shall further investigate the possibility of using GenAI in the UX research field.

*B. Limitations*

In this research, some limitations must be drawn. A severe limitation is that all data analysis is done by the researchers. Furthermore, Google Scholar was chosen as one database that has no quality control. This may result in the inclusion of irrelevant, as well as gray literature [86][87]. However, with the different scopes and stages (See Section III), we declare that we have filtered relevant records. Further databases may be investigated. However, it is questionable whether further relevant articles could be found through this, as Google Scholar is the largest database.

Moreover, we have to state the literature search conducted in July 2023. Considering the development of publications over the years, further relevant articles may be published since then. It can be also discussed whether all relevant articles were identified using the eight formulated search terms.

In addition, the Quality Assessment (See Section III-C4) must be mentioned. The record number is rather low. In contrast, the ranges between the different records concerning the quality criteria were broad. This causes the threshold to become less meaningful. Moreover, we were not able to determine a threshold for the articles regarding the theoretical foundation (See Section III-C5).

Lastly, it must be declared that all data analysis was performed by the researchers. Finally, we want to provide insights into future research.

## VI. Outlook and Future Research

To conclude, all research questions were answered, and implications were drawn. Regarding the respective implications, three research gaps in the domain UX of AR were identified. This leads us to the limitations and the outlook of this SLR. Lastly, we want to derive a future research agenda based on the results of this study.

This SLR provides a comprehensive overview of AR's UX. We presented the current state of research and outlined research gaps within this field. Overall, we want to emphasize the lack of approaches within the domain of Training and Education. It is important to gather insights into the UX of the AR applications to improve both the system and the specific experience. Therefore, it is important to develop and apply suitable measurement methods. Based on this, the following aspects for future research can be concluded:

It is essential to develop and validate suitable models and frameworks that incorporate both the system characteristics of AR and UX factors. Based on such models, suitable measurement methods and metrics can be derived.

Besides this, the learning effects of the AR application should be investigated to understand the benefit of this technology in Training and Education. Bringing both together and investigating the relationship between the UX and the learning effects could be essential for designing and developing innovative teaching and learning applications. Lastly, the rapid development of GenAI must be taken into account. Applying GenAI in UX research can enhance the research process and provide benefits for researchers. These aspects should be considered in future research.

APPENDIX

TABLE V. 71 IDENTIFIED RECORDS.

| Author | Year | Source |
|---|---|---|
| Olsson, Thomas et al. | 2011 | [74] |
| Kerr, Steven J. et al. | 2011 | [88] |
| Redondo, Ernesto et al. | 2011 | [48] |
| Pribeanu, Costin; Iordache, Dragos Daniel | 2011 | [47] |
| Dhir, Amandeep; Al-kahtani, Mohammed | 2013 | [89] |
| Rehrl, Karl et al. | 2014 | [90] |
| Irshad, Shafaq; Rambli, Dayang Rohaya Bt Awang | 2014 | [91] |
| Li, Xiao; Xu, Bo | 2014 | [92] |
| Okimoto, Maria Lucia L. R. et al. | 2015 | [38] |
| Higgett, Nick et al. | 2015 | [93] |
| Scavo, Giuseppe; Wild, Fridolin | 2015 | [39] |
| Kamilakis, Manousos et al. | 2016 | [94] |
| Irshad, Shafaq; Rambli, Dayang Rohaya Awang | 2016 | [95] |
| Seppälä, Kaapo et al. | 2016 | [96] |
| Aromaa, Susanna et al. | 2017 | [40] |
| Rashid, Zulqarnain; Pous, Rafael | 2017 | [97] |
| Dirin, Amir; Laine, Teemu H. | 2018 | [98] |
| Cheng, Kun-Hung | 2018 | [99] |
| Han, Dai-In et al. | 2018 | [62] |
| Helin, Kaj et al. | 2018 | [41] |
| Hammady, Ramy et al. | 2018 | [100] |
| Irshad, Shafaq et al. | 2018 | [61] |
| Sekhavat, Yoones A.; Parsons, Jeffrey | 2018 | [101] |
| Heo et al. | 2018 | [42] |
| Alavesa, Paula; Pakanen, Minna | 2018 | [102] |
| Jakobsen, Christian L. et al. | 2018 | [103] |
| Stumpp, Stefan et al. | 2019 | [104] |
| Kim-Berman, Hera et al. | 2019 | [51] |
| DAVIDAVIČIENĖ, Vida et al. | 2019 | [105] |
| Marques, Bernardo; Carvalho, Raphael | 2019 | [106] |
| Andri, Chairil; Alkawaz, Mohammed Hazim | 2019 | [107] |
| Sarkar, Pratiti; Pillai, Jayesh S. | 2019 | [49] |
| Kazanidis, Ioannis; Pellas, Nikolaos | 2019 | [50] |
| Satti, Fahad Ahmed et al. | 2019 | [108] |
| Cauchi, Mattea; Scerri, Daren | 2019 | [109] |
| Smaragdina, Azhar Ahmad et al. | 2019 | [52] |
| Vrellis, Ioannis et al. | 2020 | [53] |
| Irshad, Shafaq et al. | 2020 | [110] |
| Brata, Komang Candra; Liang, Deron | 2020 | [111] |
| Wang, Lei; Lv, Meiyu | 2020 | [112] |
| Harrington, Maria C. R. | 2020 | [113] |
| Mikropoulos, Tassos A. et al. | 2020 | [114] |
| García Muñzer, M. | 2020 | [115] |
| Thomaschewski, Lisa et al. | 2020 | [43] |
| Thomaschewski, Lisa et al. | 2020 | [44] |
| Domínguez Alfaro, Jessica Lizeth; Van Puyvelde, Peter | 2021 | [54] |
| DAVIDAVIČIENĖ, Vida et al. | 2021 | [116] |
| Navarro, Isidro et al. | 2021 | [117] |
| Jang, Yeonju; Park, Eunil | 2021 | [118] |
| Alenljung, Zackarias; Lindblom, Jessica | 2021 | [45] |
| Balani, Manisha Suresh; Tümler, Johannes | 2021 | [119] |
| Verhulst, Isabelle; Woods, Andy | 2021 | [120] |
| Barros et al. | 2021 | [121] |
| Syahidi, Aulia Akhrian; Tolle, Hermann | 2021 | [57] |
| Kandil, Ayman et al. | 2021 | [122] |
| Ku, Gordon Chih-Ming; Shang, I-Wie | 2021 | [123] |
| Pamparau, Christian; Vatavu, Radu-Daniel | 2022 | [124] |
| Guevara Aparicio, Ricardo Alfredo et al. | 2022 | [125] |
| Alvarez, Marina; Toet, Alexander | 2022 | [126] |
| Sudipa, Gede Iwan et al. | 2022 | [127] |
| Karimah, Hasna et al. | 2022 | [128] |
| Santana, Ronny; Rossi, Gustavo | 2022 | [56] |
| Luo, Yan et al. | 2022 | [129] |
| Liu, Enrui et al. | 2022 | [55] |
| Xue, Liangchao; Parker, Christopher J. | 2022 | [130] |
| De Paolis, Lucio Tommaso et al. | 2022 | [131] |
| Wolf, Julian et al. | 2023 | [132] |
| Hu, Shan | 2023 | [133] |
| Refae, Sema et al. | 2023 | [134] |
| Dag, Kazim et al. | 2023 | [135] |
| Gan, Quehong; Liu, Zhen | 2023 | [136] |

TABLE VI. 18 RECORDS IN THE FIELD OF TRAINING AND EDUCATION.

| Author | Focus | Year | Source |
|---|---|---|---|
| Okimoto et al. | Corporate Training | 2015 | [38] |
| Scavo and Wild | Corporate Training | 2015 | [39] |
| Aromaa et al. | Corporate Training | 2017 | [40] |
| Helin et al. | Corporate Training | 2018 | [41] |
| Heo et al. | Corporate Training | 2018 | [42] |
| Thomaschewski et al. | Corporate Training | 2020 | [43] |
| Thomaschewski et al. | Corporate Training | 2021 | [44] |
| Alenljung and Lindblom | Corporate Training | 2021 | [45] |
| Pribeanu and Iordache | Academic Teaching | 2010 | [47] |
| Redondo et al. | Academic Teaching | 2011 | [48] |
| Sarkar and Pillai | Academic Teaching | 2019 | [49] |
| Kazanidis and Pellas | Academic Teaching | 2019 | [50] |
| Kim-Berman et al. | Academic Teaching | 2019 | [51] |
| Smaragdina et al. | Academic Teaching | 2019 | [52] |
| Vrellis et al. | Academic Teaching | 2020 | [53] |
| Domínguez Alfaro and Van Puyvelde | Academic Teaching | 2021 | [54] |
| Liu et al. | Academic Teaching | 2022 | [55] |
| Santana and Rossi | Academic Teaching | 2022 | [56] |

TABLE VII. 11 IDENTIFIED RECORDS REGARDING MODELS, FRAMEWORKS, AND REVIEWS.

| Author | Focus | Year | Source |
|---|---|---|---|
| Perritaz et al. | Model/Framework | 2009 | [68] |
| Ritsos et al. | Model/Framework | 2011 | [58] |
| Olsson | Model/Framework | 2014 | [66] |
| Irshad, Shafaq; Rambli, Dayang Rohaya Awang | Model/Framework | 2015 | [59] |
| Irshad, Shafaq; Rambli, Dayang Rohaya Awang | Model/Framework | 2016 | [60] |
| Irshad, Shafaq; Rambli, Dayang Rohaya Awang | Model/Framework | 2018 | [64] |
| Ahmad Nik Azlina et al. | Model/Framework | 2023 | [63] |
| Olsson, Thomas | Review | 2013 | [73] |
| Irshad, Shafaq; Rambli, Dayang Rohaya Awang | Review | 2014 | [70] |
| Irshad, Shafaq; Rambli, Dayang Rohaya Awang | Review | 2017 | [72] |
| Arifin, Yulyani et al. | Review | 2018 | [69] |
| Ghazwani, Yahya; Smith, Shamus | Review | 2020 | [71] |
| Law, Effie Lai-Chong; Heintz, Matthias | Review | 2021 | [6] |